\documentstyle[pra,aps,prb,multicol,eqsecnum,epsf]{revtex}

\begin{document}

\title{Correlation length-exponent relation for the two-dimensional
random Ising model}

\author{P\'eter Lajk\'o$^{1}$ and Ferenc Igl\'oi$^{2,1}$}

\address{
$^1$ Institute for Theoretical Physics,
Szeged University, H-6720 Szeged, Hungary\\
$^2$ Research Institute for Solid State Physics and Optics, 
H-1525 Budapest, P.O.Box 49, Hungary\\
}
\maketitle

\begin{abstract}
We consider the two-dimensional (2d) random Ising model on a diagonal strip of the
square lattice, where the bonds take two values, $J_1>J_2$, with equal
probability. Using an iterative method, based on a successive application
of the star-triangle transformation, we have determined at the bulk
critical temperature the correlation
length along the strip, $\xi_L$, for different widths of the strip, $L \le 21$.
The ratio of the two lengths, $\xi_L/L=A$, is found to approach the universal
value, $A=2/\pi$ for large $L$, independent of the dilution parameter, $J_1/J_2$.
With our method we have demonstrated with high numerical precision, that the
surface correlation function of the 2d dilute Ising model is self-averaging,
in the critical point conformally coovariant and the corresponding decay
exponent is $\eta_{\parallel}=1$.
\end{abstract}


      \newcommand{\bc}{\begin{center}}
\newcommand{\ec}{\end{center}}
\newcommand{\be}{\begin{equation}}
\newcommand{\ee}{\end{equation}}
\newcommand{\beqn}{\begin{eqnarray}}
\newcommand{\eeqn}{\end{eqnarray}}

\begin{multicols}{2}
\narrowtext
\section{Introduction}
In the presence of quenched, i.e. time independent disorder one generally
considers different random samples
and the physical observables are characterized by their distribution
and ($n$-th) moments. (For $n=1$ and $n=0$ we have the {\it average} and
{\it typical} value, respectively.) The extensive quantities,
which are connected to the free-energy and its derivatives, have normal
distribution, thus in a single sample one measures their {\it average} value
with probability one in the thermodynamic limit. These quantities are
called {\it self-averaging}. There are, however, other observables,
typically correlation functions, which are broadly distributed and the
typical (or most probable) value is different from the average value,
even in the thermodynamic limit.

Such type of phenomena takes place in disordered quantum systems\cite{qsg}, where the typical
and average behavior of correlations and critical singularities are even
qualitatively different. As known by exact results\cite{fisher,bigpaper},
renormalization group\cite{2dRG}
and numerical calculations\cite{2dmc} in the {\it infinite randomness fixed
point} the average behaviour is dominated by the {\it rare events},
which occur with vanishing probability, whereas the typical behaviour is seen in
any large sample with probability one.

In a classical system the effect of disorder is comparatively weaker, (since in quantum
systems the disorder is strictly correlated along the (imaginary) time direction), here
the critical singularities are controlled by a {\it random fixed point}
and there are usually quantitative differences between the average and typical behaviour.

In this respect a well known example is the one-dimensional (1d) random bond Ising
model\cite{derrida},
defined by the Hamiltonian $H=-\sum_j J_j s_j s_{j+1}$. Here the spin correlation
function
\be
G(r)=\langle s_{j+r} s_j \rangle=\prod_{k=j}^{j+r-1} t_k~~,~~t_k=\tanh (J_k/k_B T)\;,
\label{1dising}
\ee
is given as a product
of random numbers and has a log-normal distribution. Consequently its most probable, or typical
value $G(r)_{\rm typ}=\exp[\ln(G(r)]_{\rm av}=[t]_{\rm av}^r$ and average
value $[G(r)]_{\rm av}$ are different, even in the thermodynamic limit. In the
following we use $\langle \dots \rangle$ and $[ \dots ]_{\rm av}$ to denote thermal and
disorder averaging, respectively.

In higher dimensional classical spin systems with random ferromagnetic couplings, such as
the random Ising and $Q$-state
Potts models, the effect of disorder is expected to be even weaker than in 1d. In calculating the
correlation function the thermal average in higher dimensions involves several random
couplings, not only those connecting directly the two points, therefore the disorder fluctuations
are smoothed down. There is a class of random systems which, in the vicinity
of their critical point, are homogeneous in macroscopic scales, thus the effect of quenched
disorder is {\it irrelevant}.  The corresponding criterion for weak randomness
due to Harris\cite{harris} requires $\alpha^{\rm pure} < 0$, where $\alpha^{\rm pure}$
is the specific heat exponent of the pure system.

For systems with $\alpha^{\rm pure}>0$ the disorder is a {\it relevant} perturbation so that the
critical properties are controlled by a (new) {\it random fixed point} in which unconventional
scaling behavior is expected.  A detailed study, both (field)-theoretical\cite{ludwig,lewis} and
numerical\cite{cj,oy,pcbi},
about the two-dimensional random $Q>2$ state Potts model has revealed that
the critical bulk spin correlation
function has {\it multifractal} behavior: the different moments of the correlation
function at the critical point decay as a power:
\be
[G^n(r)]_{\rm av}^{1/n} \sim r^{-2 x^{(n)}}\;,
\label{multifr}
\ee
with $n$ dependent decay exponents, $x^{(n)}$. We note that for conventional
scaling the $x^{(n)}$-s have no $n$-dependence. 

An important question concerning random magnetic systems is whether the critical point
correlations in Eq.({\ref{multifr}) transform coovariantly under conformal transformations\cite{Cardy}.
Although correlations in one sample are not translationally invariant the {\it average}
correlations are translationally and rotationally invariant and - it is generally believed -
they are also conformally coovariant. Indeed numerical studies in the strip and rectangle
geometry for the two-dimensional $Q>2$ state random Potts model show that
average critical correlations transform coovariantly under
conformal transformations\cite{CB}. Recently conformal properties of correlation
functions and density profiles have been used to determine the scaling dimensions
of different operators\cite{pcbi}.

The two-dimensional random Ising model with $\alpha^{\rm pure}=0$ represents the marginal
situation of the Harris criterion and detailed studies have been performed to clarify its
critical properties\cite{SST}. Disorder is predicted as a marginally irrelevant perturbation
by field theory, so that the critical singularities of the random model are characterized
by the exponents of the pure Ising model supplemented by logarithmic
corrections\cite{dotsenko,shal,ludwig}.
Numerical studies are in favour of this scenario\cite{SST,QS,SAQS,SSLI,ILSS,RAJ},
although conflicting interpretation
of the numerical results has also been suggested\cite{kuhn,kim}.

Considering the spin correlation function of the random Ising model
according to field theory the decay of the different moments of the bulk correlations
at the critical point are given by the power law of the pure model
with $x^{\rm pure}=1/8$, but the logarithmic corrections to the different moments are
$n$ dependent\cite{ludwig,DPP}. Numerically the bulk critical correlations are studied in the
infinite plane geometry by MC simulations in Ref[\onlinecite{oy}], however the possible
logarithmic corrections of the moments have not been analyzed. On the other hand
in Ref[\onlinecite{QS,Q}] the transfer matrix method is used in the strip geometry and the
decay exponent of the typical and average bulk correlations are deduced from the
assumption of conformal invariance.

In the semi-infinite geometry the critical surface correlation
function has been studied in Ref[\onlinecite{ILSS}] by the star-triangle (ST) method .
It was found that for
any dilution the numerically calculated average correlations are compatible with the form:
\be
[G_s(r)]_{\rm av} \sim r^{-1} (\ln r)^{1/2}\;.
\label{semiinf}
\ee
Thus the decay exponent of the critical surface magnetization is $\eta_{\parallel}=1$,
as for the pure system, however in the random model
there are also logarithmic corrections.

In this paper we continue to study the surface correlation function of the 2d random
Ising model. New features of our investigations are the following.

i) We considered the strip geometry, rather than the semi-infinite geometry.

ii) As a numerical method we used an iterative procedure based on the star-triangle
transformation. By this ST method a finite strip of random Ising model is formally
transformed
to a chain of Ising spins with smoothly inhomogeneous bonds. Then, using the exact
expression in Eq.(\ref{1dising}), we have calculated very accurately the correlation
length parallel to the strip, $\xi_L$, and studied its distribution and different moments.

iii) The advantage of the ST method to the transfer matrix (TM) technique, applied
previous for bulk correlations is two-fold. First, we could investigate
larger widths of the strip, going up to $L=21$, which is approximately
twice of the widths available by the TM technique\cite{Q}. The second advantage of
the ST method that one can consider correlations between two largely separated spins,
$r=O(10^3)$, which is at least one order of magnitude larger, than for the
TM method. In this way we obtained more accurate averaging and could go deeper
into the asymptotic region of the correlations.
 
iv) Finally, we calculated different moments of the correlation function and studied
the validity of the correlation length - exponent relation, as follows from the
assumption of conformal invariance.

The structure of the paper is the following. The model and the ST method to calculate
surface correlations are presented in Section 2. Our results about surface
correlations and the corresponding correlation lengths are given in Section 3.
We conclude our paper with a Discussion in the final Section.

\section{Star-triangle approach to surface correlations}

We consider the Ising model on a diagonal strip of the square lattice,
with $i=1,2,\dots,L$ columns and $j=1,2,\dots,K$ rows. At $i=1$ and $i=L$
there are two $(1,1)$ surfaces, whereas in the vertical direction with
$K \gg L$ we impose periodic boundary conditions. The nearest
neighbor spins are connected with ferromagnetic
couplings, $J_{i,j}>0$, which could
take two values, $J_1>J_2$, with equal probability. In the thermodynamic
limit $L,K \to \infty$ the model is self-dual\cite{fisch} and the self-duality point:
\be
\tanh (J_1/k_B T)=\exp(-2 J_2/k_B T)\;,
\label{criticalpoint}
\ee
corresponds to the critical point, since according to
numerical studies there is one
phase-transition in the system. The degree of dilution can be varied
by changing the ratio of the strong and weak couplings, $\rho=J_1/J_2$.
At $\rho=1$ one recovers the perfect Ising model, whereas for $\rho \to
\infty$ we are in the percolation limit, where $T_c=0$.

For a given distribution of the couplings correlations between two
surface spins, $G_s(r)=\langle s_{1,j+r} s_{1,j} \rangle$, can be conveniently
calculated by the star-triangle method. The ST method was introduced
by Hilhorst and van Leeuwen\cite{HvL} and later by others\cite{others} to study the surface
critical properties of triangular lattice Ising models with a layered
structure. Recently, the method has been generalized for arbitrary
distribution of the couplings and applied for the random semi-infinite
Ising model in Ref[\onlinecite{ILSS}], hereafter referred to as Paper I.
In the following we recapitulate the method for the
{\it strip geometry} with free boundary conditions at the two edges
of the strip.

\input{epsf.sty}
    \begin{figure}
\hspace{15mm} \epsfxsize= 4,5cm \epsfbox{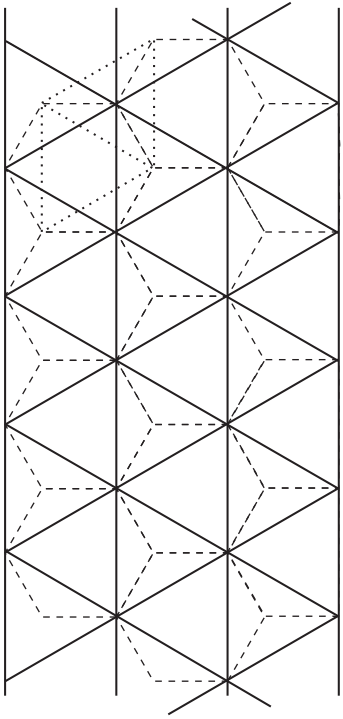}
\vspace{5mm}
\caption{Strip of triangular lattice with $L=4$ layers. The hexagonal lattice,
obtained by the ST transformation is denoted by dashed lines, the triangles of
the new triangular lattice are shown by dotted lines. In a diagonal square
lattice the vertical couplings, which are zero in the original model, are
generated during the iteration process.}
\label{fig:1}
\end{figure}

Central to the method is the ST transformation by which one replaces
all right-pointing triangles of the strip by a star, which yields a
hexagonal lattice of spins, denoted by dashed lines in Fig. 1. In the
second step of the mapping the left pointing stars of the hexagonal
lattice are replaced by triangles resulting in a new triangular
lattice, which is denoted by dotted lines in Fig. 1. Iterating the
procedure a sequence of triangular Ising models is generated
($m=0,1,2,\dots$) from the original model with $m=0$. Neither the
width of the strip nor the number of spins is changed under the transformation.

As seen in Fig. 1 the surface spins of the $m$-th and the $(m+1)$-th models
are connected by the surface couplings of the intermediate hexagonal
lattice and, as shown in Paper I, there is an explicit relation between the
thermal average of the surface spins in the two models. Then, from the fact
that the surface magnetization of a finite strip with free boundary conditions
is vanishing, follows that the surface couplings of the hexagonal lattice
goes to zero as $m \to \infty$. As a consequence the surface spins of the
triangular lattice decouple asymptotically from the rest of the system.
The surface correlation functions of the $m$-th and $(m+1)$-th triangular models
are similarly related and one can use the results of Paper I to calculate
the surface correlation length from this relation. One can, however, proceed
in a simpler way noticing that the
surface spin correlation function stays asymptotically invariant under the
mapping. Then $G_s(r)$ in the original model can be expressed in the form of the
one-dimensional Ising model in Eq.(\ref{1dising}) replacing $J_j$ by
the asymptotic value of the surface coupling $J^{(s)}_j(m) \equiv J_{1,j}(m)$.

For a given strip of width $L$ the surface correlations show an asymptotic
exponential decay $G_s(r) \sim \exp(-r/\xi_L)$, $r \gg L$, where the correlation
length, $\xi_L$, is approximated by
\vspace{-15mm}
\begin{figure}
\hspace{-10mm} \epsfxsize=130mm \epsfbox{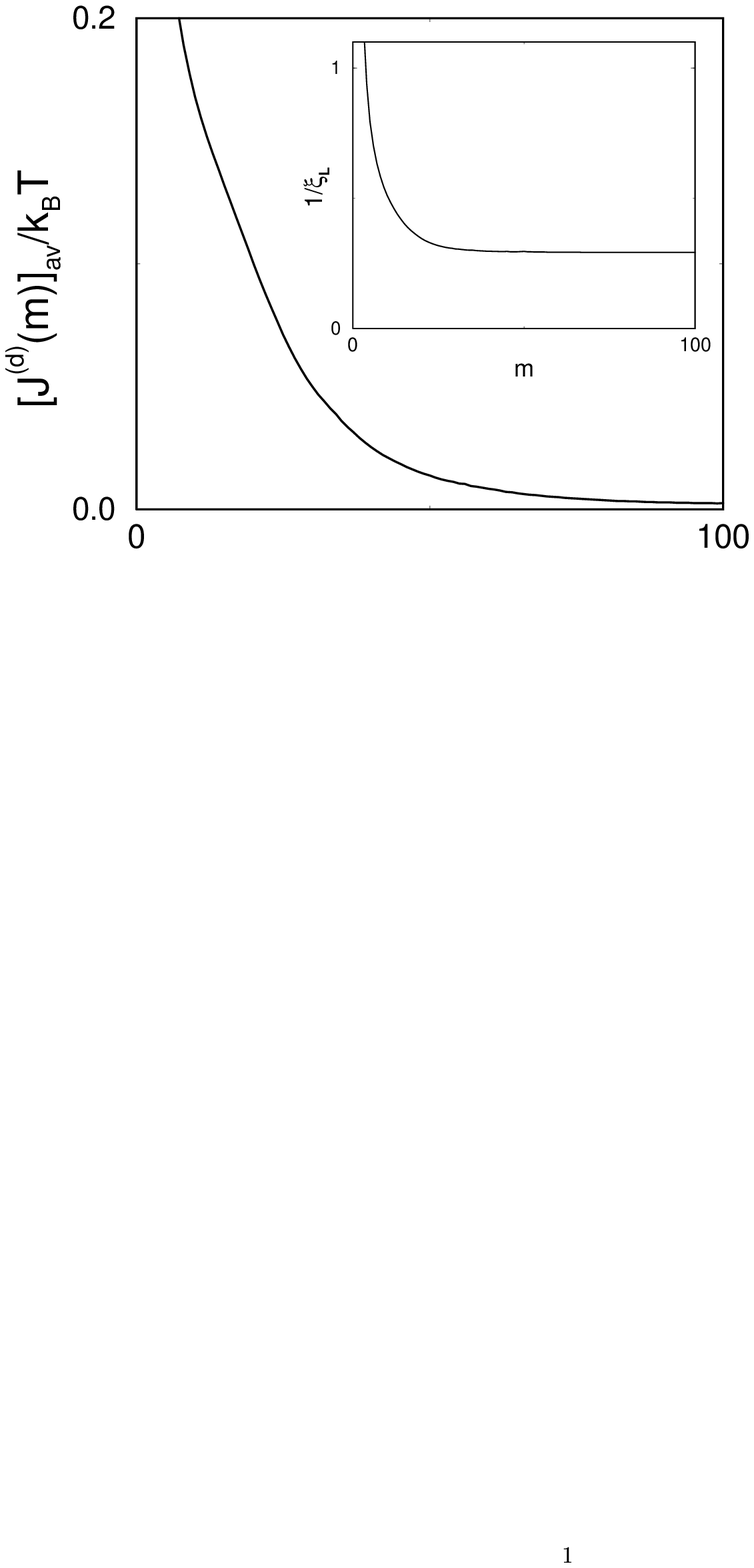}
\vspace{-95mm}
\caption{The average surface diagonal coupling, connecting the surface spins to the
rest of the system, as a function of the number of iteration, $m$. The calculation is
performed on a given random sample with dilution $\rho=4$ and on a strip of size $L=11$
and $K=1024$.
Inset: The approximate inverse correlation length, defined in Eq.(2.2), as a function
of the iteration parameter.}
\label{fig:2}
\end{figure}
\be
{1 \over \xi_L(m)}=-{1 \over r} \ln\left( \prod_{k=j}^{j+r-1} \tanh (J^{(s)}_k(m)/k_B T)
\right)\;,
\label{xiLm}
\ee
and $\lim_{m \to \infty} \xi_L(m)=\xi_L$. Averaging over different
disorder realizations one obtains the {\it typical} correlation length:
\be
\xi_L^{\rm typ}=[\xi_L]_{\rm av}\;.
\label{xityp}
\ee

In physical applications one should average the $n$-th power of the
correlation function the asymptotic behavior of which,
$[G_s(r)^n]_{\rm av}^{1/n} \sim \exp(-r/\xi_L^{(n)})$,
defines the corresponding correlation length, $\xi_L^{(n)}$. As already
mentioned for $n=1$ and $n=0$ we obtain the average and typical
correlation lengths, respectively.
From the different moments of the distribution of the inverse
correlation length
$p(1/\xi_L)$ one obtains $\xi_L^{(n)}$ in a cumulant expansion:
\be
{1 \over \xi_L^{(n)}}={1 \over \xi_L^{\rm typ}}-{1 \over 2} nr\left[
\left({1 \over \xi_L}-{1 \over \xi_L^{\rm typ}}\right)^2 \right]_{\rm av}
+\dots\;.
\label{cum}
\ee
For isotropic systems the correlation length parallel to the strip, $\xi_L^{(n)}$
and the width of the system, $L$, are asymptotically proportional and for
{\it conformally invariant} systems their ratio takes the universal value\cite{Cardy}:
\be
{\xi_L^{(n)} \over L}={1 \over \pi x_s^{(n)}}\;.
\label{corr-exp}
\ee
Here $x_s^{(n)}$ is the anomalous dimension of the surface magnetization,
defined through the asymptotic decay of the critical surface correlation in the semi-infinite
geometry, $[G_s(r)^n]_{\rm av}^{1/n} \sim r^{-2x_s^{(n)}}$. Thus $x_s^{(n)}$ is the surface
counterpart of $x^{(n)}$ in Eq.(\ref{multifr}) and satisfies the scaling relation
$\eta_{\parallel}^{(n)}=2 x_s^{(n)}$.

\section{Results}

We studied the spin correlations of the random Ising model on the $(1,1)$ surface
of the square lattice by the ST method. Evidently the original model with $m=0$
can be considered as a special triangular lattice model with vanishing
vertical bonds. During iteration, however, non-zero vertical couplings are
generated so that also the surface couplings $J^{(s)}_j(m)$ become non-zero.
In the actual calculations we considered strips of width $L=2l+1$ up to
$L=21$\cite{width}, whereas for the length of the strip, $K$, the condition
$K \gg L$ is always satisfied. Typically we took $K=1024$ and checked that
the numerical results are insensitive on the variation of $K$ in this region.

\begin{figure}
\hspace{0mm} \epsfxsize=\columnwidth \epsfbox{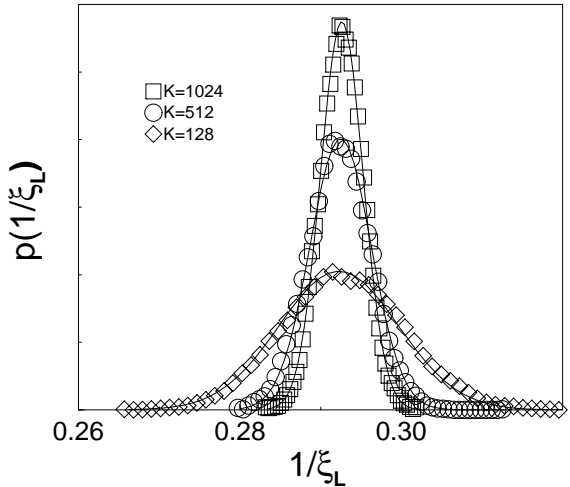}
\vspace{5mm}
\caption{ Probability distribution of the inverse of the correlation length for $L=11$
and for different lengths of the strip, $K$. We studied 20000 samples for  each lengths
at a dilution $\rho=4$. The full lines represent the Gaussian approximations to the
distributions.}   
\end{figure}

As indicated in the previous chapter, under iteration the surface spins asymptotically
decouple from the rest of the system and the surface correlations have one-dimensional
character. For an illustration we have calculated the average value of the
first diagonal coupling, $[J^{(d)}(m)]_{\rm av}$, connecting the first and second
line of spins, as a function of the iteration, $m$. It is given in
Fig. 2 together with the correlation length, $\xi_L(m)$, as defined in Eq. (\ref{xiLm})
between two spins of maximal distance, $r=K/2$. As seen in Fig. 2 both
$[J^{(d)}(m)]_{\rm av}$ and $\xi_L(m)$ approach their limiting values rapidly,
exponentially with $m$. Analyzing the iteration equations in Paper I one can
show that $m \sim L^2$ iteration steps are needed to reach the asymptotic region,
which is indeed verified numerically.

For a given dilution, $\rho$, the correlation length, $\xi_L$, shows variation from sample
to sample. The distribution
of the inverse correlation length, $p(1/\xi_L)$, obtained over 20000 samples
 is shown in Fig. 3 for different lengths
of the strip, $K$. As seen in Fig. 3 the average value of $\xi_L$, defining the
{\it typical} correlation length in Eq.(\ref{xityp}) is independent of $K$, whereas
the width of the distribution is decreasing with the length of the strip as
$1/\sqrt{K}$. This observation is in agreement with the cumulant expansion in
Eq.(\ref{cum}) and with the fact that $\xi_L^{(n)}$ is asymptotically independent of $K$.
The distribution $p(1/\xi_L)$ is found approximately Gaussian, however for finite
strips there is always some deviation from the normal distribution.

\begin{figure}
\hspace{0mm} \epsfxsize=\columnwidth \epsfbox{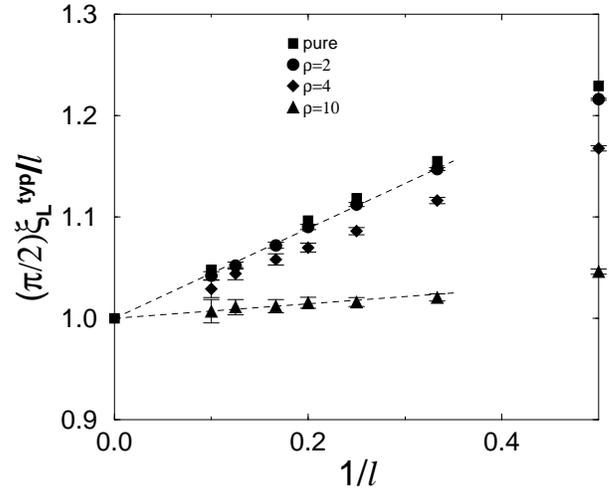}
\vspace{5mm}
\caption{Ratio of the typical correlation length, $\xi_L^{\rm typ}$, and the with of the
strip, measured in units $l=(L-1)/2$ for different dilutions. The dashed lines are guide
to the eyes, representing the finite-size corrections as $O(1/l)$.}
\end{figure}

Next, we are going to study the $L$ dependence of the typical correlation length,
$\xi_L^{\rm typ}$, for different values of the dilution $\rho=1,2,4$ and $10$. 
First we note that in order to use the same lattice units in the vertical and
horizontal directions one should replace $L=2l+1$ by $l$.
In
Fig. 4 the ratio $(\pi/2) \xi_L^{\rm typ}/l$ is plotted against $1/l$. As seen
in Fig. 4 in the range of dilution we worked, the correlation length is monotonically
decreasing with $\rho$, whereas there is an approximate linear $1/l$ correction
to the ratio for all values of $\rho$. The asymptotic value of the ratio is found
dilution independent, we estimated as
\be
\lim_{L \to \infty} {\pi \over 2} {\xi_L^{\rm typ} \over l}=1 \pm 0.003\;.
\label{ratiotyp}
\ee

We have also studied the different moments of the correlation function and calculated
the corresponding average correlation length, $\xi_L^{(n)}$. The ratio
$(\pi/2)\xi_L^{(n)}/l$ is found to have a strong size dependence, much stronger
than for the typical correlation length. In this case the finite-size corrections
are approximately logarithmic, as can be seen in Fig. 5. For any finite $L$ the
corrections are increasing with the moment, $n$, however the asymptotic value
of the ratio tends to the same universal value
\be
\lim_{L \to \infty} {\pi \over 2} {\xi_L^{(n)} \over l}=1 \pm 0.03\;.
\label{ratioav}
\ee
as for the typical ratio in Eq.(\ref{ratiotyp}). We note that the error of the estimate
in Eq.(\ref{ratioav}) is increasing with $n$.
\begin{figure}
\hspace{0mm} \epsfxsize=\columnwidth \epsfbox{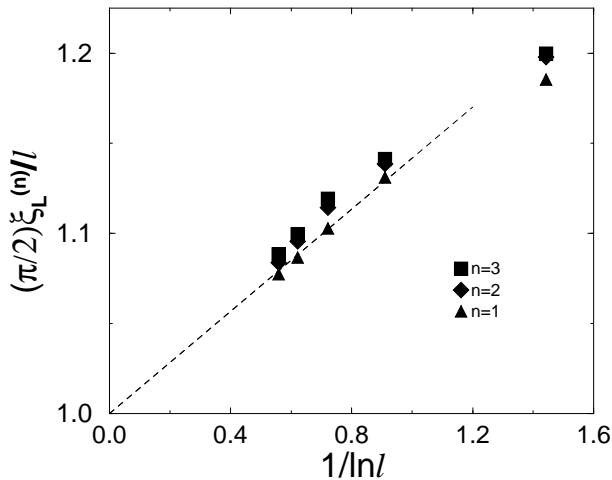}
\vspace{5mm}
\caption{Ratio of the correlation length corresponding to the $n$-th moment of the correlation
function, $\xi_L^{(n)}$, and the width of the strip in units of $l=(L-1)/2$ at a
dilution $\rho=4$. The dashed line is guide to the eyes, representing logarithmic
finite-size corrections as $O(1/\ln l)$.}
\end{figure}

At this point we compare our results in Eqs. (\ref{ratiotyp}) and (\ref{ratioav}) to
that obtained in the semi-infinite geometry in Eq.(\ref{semiinf}) and comment about
the validity of the correlation length-exponent relation in Eq.(\ref{corr-exp}).
The universal ratio found in Eqs. (\ref{ratiotyp}) and (\ref{ratioav}) according
to the correlation length-exponent relation corresponds to a typical and average
surface magnetization decay exponent of $\eta_{\parallel}=1$ in complete agreement
with the result in the semi-infinite geometry in Eq.(\ref{semiinf}). Furthermore,
for the average correlations we observed logarithmic corrections in both geometries.
Thus we can conclude that the correlation length-exponent relation is valid for the
typical and average surface correlations of the random 2d Ising model, thus they
transform coovariantly under conformal transformations.

\section{Discussion}

In the previous section the surface correlation function of the
2d random Ising model is studied in the strip geometry. In particular
we have calculated the ratio of the correlation length, $\xi_L^{(n)}$,
obtained from the average of the $n$-th moment of the surface
correlation function, and the width, $L$, of the system. We found
that asymptotically this ratio goes to a universal value, irrespective
of the degree of dilution, $\rho$, and the value of the moment, $n$. For
typical correlations the correction terms are of $O(1/L)$, whereas
for the average, $n=1$, and for the higher moments, $n>1$. The finite-size
corrections are logarithmic, they are in the form of $O(1/\ln L)$.

Several qualitative features of the above results can be understood by analyzing the
ST iteration procedure. As mentioned before the surface spins are asymptotically
decoupled after $m \sim L^2$ iteration steps, when the expression of the
new surface couplings,
$J_k^{(s)}(m)$, contains a set of the original couplings, $J_{i,j}$, taken
from a region of $1 \le i \le L$ and $k-L < j < k+L$. Consequently the
$J_k^{(s)}(m)$ are correlated for short distances, but they are
practically independent between two sites which are separated by at least a
distance of $O(L)$. Now we can use an approximate coarse-grained
description: introduce block-spin variables to replace each $O(L)$
number of surface spins which are connected by correlated couplings. The new
couplings between the block-spins, $J^{(B)}_{k'}$, are approximately independent
random variables and they satisfy $\tanh (J^{(B)}_{k'}/k_B T)=O(1)$, since
with the original couplings $\tanh(J^{(s)}_k/k_B T)=O(1/L)$.
Then from Eq.(\ref{xiLm}) follows that the correlation length of the
system is self-averaging, in any sample the expression in Eq.(\ref{xiLm})
in the thermodynamic limit goes to the average correlation length,
$[\xi_L]_{\rm av}$, with probability one. Evaluating the variance of
$1/\xi_L$ in the coarse-grained picture one gets:
\be
\left[\left({1 \over \xi_L} -{1 \over \xi_L^{\rm typ}}\right)^2\right]_{\rm av}
\simeq{\beta \over r L}\;,
\label{variance}
\ee
thus the first correction term in the cumulant expansion in Eq.(\ref{cum})
is of $O(1/L)$, as it should be. The eventual correlations between the
block-spin couplings will tend to reduce the value of $\beta$ in
Eq.(\ref{variance}). According to our numerical studies $\beta \sim 1/\ln L$
and this effect is the source of the logarithmic corrections in the
random Ising model.

The results in the semi-infinite geometry and in the strip geometry are
in complete correspondence. Comparing the conformal result in
Eq.(\ref{corr-exp}) with the numerical estimates in Eqs.(\ref{ratiotyp})
and (\ref{ratioav}) one obtains the following conclusions.

i) The correlation length-exponent relation is valid for the random
Ising model, thus the (surface) correlations of the system are
conformally coovariant.

ii) For the typical and average correlations at the critical point
the decay is given by the same exponent, which does not depend on the
degree of dilution. Consequently there is no multifractal behavior
for the critical correlations of the model.

iii) The typical surface correlations are free of logarithmic
corrections, whereas the average correlations and the higher moments
are subject of logarithmic corrections, the strength of those is increased
with $n$.

iv) Finally, our numerical results
give strong and accurate numerical support to the field-theoretical
conjecture that the random and pure Ising models in 2d belong to
the same universality class.

Acknowledgment: 
F.I. is indebted to B. Berche, H.J. Hilhorst and L. Turban for useful discussions.
This work has been supported by the Hungarian National Research Fund
under grant No OTKA TO23642, TO25139, F/7/026004, M 028418 and by the Ministery of
Education under grant No. FKFP 0596/1999. L.P. thanks the Soros Foundation,
Budapest, for a traveling grant.

}
\end{multicols}

\begin{thebibliography}{99}
\bibitem{qsg} 
        See H. Rieger and A. P Young, in {\it Complex Behavior
        of Glassy Systems}, ed.\ M. Rubi and C. Perez-Vicente, Lecture
        Notes in Physics {\bf 492}, p.\ 256, Springer-Verlag,
        Heidelberg, 1997, for a review on the Ising quantum spin glass
        in a transverse field.

\bibitem{fisher}
        D.S. Fisher, Phys. Rev. Lett. {\bf 69}, 534 (1992); 
        Phys. Rev. B {\bf 51}, 6411 (1995).

\bibitem{bigpaper}
        F. Igl\'oi and H.\ Rieger, Phys. Rev. B {\bf 57}, 11404 (1998);
	F. Igl\'oi and H. Rieger, Phys. Rev. E {\bf 58}, 4238 (1998)

\bibitem{2dRG}
	O. Motrunich, S.-C. Mau, D.A. Huse and D.S. Fisher, cond-mat/9906322

\bibitem{2dmc}
	C. Pich, A.P. Young, H. Rieger and N. Kawashima, Phys. Rev. Lett.
	{\bf 81}, 5916 (1998)

\bibitem{derrida}
	B. Derrida, Phys. Rep. {\bf 103}, 29 (1994)

\bibitem{harris}
	A.B. Harris, J. Phys. C {\bf 7}, 1671 (1974)

\bibitem{ludwig}
	A.W.W. Ludwig, Nucl. Phys. B {\bf 330}, 639 (1990).
\bibitem{lewis}
	M.A. Lewis, Europhys. Lett. {\bf 43}, 189 (1998).
\bibitem{cj}
	J.L. Cardy and J.L. Jacobsen, Phys. Rev. Lett. {\bf 79}, 4063 (1997);
	J.L. Jacobsen and J.L. Cardy, Nucl. Phys. B {\bf 515}, 701 (1998).
\bibitem{oy}
	T. Olson and A.P. Young, cond-mat/9903068.
\bibitem{pcbi}
	G. Pal\'agyi, C. Chatelain, B. Berche and F. Igl\'oi, Eur. Phys. J. B (in print),
	cond-mat/9906067.
\bibitem{Cardy}
	For a review see: J.L. Cardy, in {\it Phase Transitions and Critical
	Phenomena} Vol. 11 eds. C. Domb and J.L. Lebowitz (New York: Academic) (1987)
\bibitem{CB}
	C. Chatelain and B. Berche, Phys. Rev. E {\bf 58}, R6899 (1998); cond-mat/9902212.
\bibitem{SST}
	See W. Selke, L.N. Shchur and A.L. Talapov, {\it Annual Reviews of Computational
	Physics}, vol 1, ed D. Stauffer (Singapore: World Scientific) p. 17 (1994) for a
	review on dilute Ising models.
\bibitem{dotsenko}
	Vik. S. Dotsenko and Vl. S. Dotsenko, Adv. Phys. {\bf 32}, 129 (1983);
	V. Dotsenko, Usp. Fiz. Nau. {\bf 165}, 481 (1995).
\bibitem{shal}
	B.N. Shalaev, Phys. Rep. {\bf 237}, 129 (1994).
\bibitem{QS}
	S.L.A. de Queiroz and R.B. Stinchcombe, Phys. Rev. E {\bf 54}, 190 (1996).
\bibitem{SAQS}
	D. Stauffer, F.D.A. Aar\~ao Reis, S.L.A. de Queiroz and R.R. dos Santos,
	Int. J. Mod. Phys. C {\bf 8}, 1209 (1997).
\bibitem{SSLI}
	W. Selke, F. Szalma, P. Lajk\'o and F. Igl\'oi, J. Stat. Phys. {\bf 89},
	1079 (1997); F. Szalma and F. Igl\'oi, J. Stat. Phys. {\bf 95}, 763 (1999).
\bibitem{ILSS}
	F. Igl\'oi, P. Lajk\'o, W. Selke and F. Szalma, J. Phys. A {\bf 31}, 2801 (1998).
\bibitem{RAJ}
	A. Roder, J. Adler and W. Janke, cond-mat/9905255
\bibitem{kuhn}
	R. K\"uhn, Phys. Rev. Lett. {\bf 73}, 2268 (1994)
\bibitem{kim}
	J-K. Kim and A. Patrascioiu, Phys. Rev. Lett. {\bf 72}, 2785 (1994);
	J-K. Kim, cond-mat/9905202.
\bibitem{DPP}
	V. Dotsenko, M. Picco and P. Pujol, Nucl. Phys. B {\bf 455} [FS] 701 (1995)
\bibitem{Q}
	S.L.A. de Queiroz, J. Phys. A {\bf 30}, L443 (1997)
\bibitem{fisch}
	R. Fisch, J. Stat. Phys. {\bf 18}, 111 (1978)
\bibitem{HvL}
	H.J. Hilhorst and J.M.J. van Leeuwen, Phys. Rev. Lett. {\bf 47}, 1188 (1981)
\bibitem{others}
	T.W. Burkhardt, I. Guim, H.J. Hilhorst and J.M.J. van Leeuwen, Phys. Rev. B
	{\bf 30}, 1486 (1984); F. Igl\'oi and P. Lajk\'o, J. Phys. A {\bf 29}, 4803 (1996);
	T.W. Burkhardt and I. Guim, Physica A {\bf 251}, 12 (1998).
\bibitem{width}
	In principle one could study much larger strips, for example in the
	perfect system one can easily go up to $L=1000$. For the random
	system, however, there occur regions in the sample with many weak
	(or strong) couplings and therefore the iteration procedure could become	numerically instable with increasing width of the system.

\end{thebibliography}
\end{document}